# Strongly enhanced topological quantum phases in dual-surface AlO$_x$-encapsulated MnBi$_2$Te$_4$

Zichen Lian[1†], Yongqian Wang[2,3†], Yongchao Wang[1†], Liangcai Xu[1], Jinsong Zhang[1,4], Chang Liu[2,3*], Yayu Wang[1,4,5*]

[1]*State Key Laboratory of Low Dimensional Quantum Physics, Department of Physics, Tsinghua University, Beijing 100084, P. R. China*

[2]*Beijing Key Laboratory of Opto-electronic Functional Materials & Micro-Nano Devices, Department of Physics, Renmin University of China, 100872 Beijing, China*

[3]*Key Laboratory of Quantum State Construction and Manipulation (Ministry of Education), Renmin University of China, Beijing 100872, China*

[4]*Hefei National Laboratory, Hefei 230088, China*

[5]*New Cornerstone Science Laboratory, Frontier Science Center for Quantum Information, Beijing 100084, P. R. China*

[†] *These authors contributed equally to this work.*

\* Emails: liuchang_phy@ruc.edu.cn; yayuwang@tsinghua.edu.cn

**ABSTRACT**. The topological quantum phases in antiferromagnetic topological insulator MnBi$_2$Te$_4$ hold promise for next-generation spintronics, but their experimental realization has been constrained by challenges in preparing high-quality devices. In this work, we report a new wax-assisted exfoliation and transfer method that enables the fabrication of MnBi$_2$Te$_4$ heterostructures with both surfaces encapsulated by AlO$_x$. This strategy strongly enhances the topological quantum phases in MnBi$_2$Te$_4$ flakes. We observe the robust axion insulator state in even-layer device with wide zero Hall plateau and high longitudinal resistivity, and the quantum anomalous Hall effect in odd-layer device with large hysteresis and sharp plateau transition. These results demonstrate that the combination of wax exfoliation and AlO$_x$ encapsulation provides great potentials for exploring novel topological quantum phenomena and potential applications in MnBi$_2$Te$_4$ and other two-dimensional materials

## I. INTRODUCTION.

The intrinsic magnetic topological insulator MnBi$_2$Te$_4$ (MBT) uniquely combines nontrivial band topology with antiferromagnetic (AFM) order [1–3]. Its bulk crystal consists of Te–Bi–Te–Mn–Te–Bi–Te septuple layers (SLs) stacked along the z-direction with weak van der Waals bonding [4]. The magnetic moments of Mn ions have ferromagnetic (FM) alignment within each SL and AFM order between neighboring SLs. When exfoliated to few-SL flakes, the magnetic properties exhibit intriguing evolution and lead to different topological quantum phases that vary dramatically with thickness and magnetic fields [5–9]. In odd-SL MBT films, the parallel magnetizations at the top and bottom surfaces lead to the quantum anomalous Hall (QAH) phase at zero-magnetic-field with dissipationless chiral edge state [10,11]. The tunable exchange bias observed in odd-SL samples presents exciting opportunities for the development of spintronic devices [12–14]. In even-SL MBT, the antiparallel magnetizations of the two surfaces lead to the axion insulator state with a zero Hall plateau in the low-field AFM state and quantized Chern insulator phase in the high-field FM state [15]. There are also observations of the layer Hall effect [16], non-linear transport [17,18], and dynamic axion quasiparticles [19] in even-layer devices.

However, the nano-fabrication process for MBT transport devices often introduce complexities that degrade device quality, making it difficult to observe or reproduce the topological quantum effects [20,21]. Theoretical calculations have also indicated that the electronic state near the surface of MBT is susceptible to perturbations [22–24]. We noticed that contact with $AlO_x$ significantly enhances the anomalous Hall resistance of MBT at zero magnetic field [10,25–27]. Recently, we demonstrated that depositing an $AlO_x$ capping layer on the MBT top surface significantly improves the device quality, which allowed us to observe the QAH effect modulated by spin flips and flops that are unique to the AFM topological insulator [11,28]. The $AlO_x$ layer not only protects the surface during the fabrication process, but also enhances the perpendicular magnetic anisotropy (PMA) of the interface MBT layer. Given the significant role of $AlO_x$ in enhancing interface quality and magnetic properties, a natural question arises that whether having both surfaces of MBT in contact with $AlO_x$ can lead to more robust topological quantum phase. The strategy of interface engineering from single-surface to dual-surface for improving device quality has been widely demonstrated in studies on graphene devices protected by *h*-BN [29,30]. However, the fabrication of MBT devices with both surfaces encapsulated by $AlO_x$ poses substantial technical challenges. The dry-transfer methods commonly used for fabricating graphene heterostructures have seen limited application in MBT. A key challenge is the 90 °C temperature required for transferring polypropylene carbonate (PPC), which may cause changes to the chemical composition of MBT [31].

In this work, we introduce a new wax-assisted exfoliation and transfer method for fabricating MBT devices with both surfaces encapsulated by $AlO_x$ [32]. This dual-capping strategy strongly enhances the topological quantum phases in MBT. We observe a well-developed axion insulator state in 6-SL device, and the quantum anomalous Hall effect in 7-SL device with larger hysteresis than previously reported. These results demonstrate that the combination of wax exfoliation and $AlO_x$ encapsulation provides great potentials for exploring novel topological quantum phenomena and potential applications in MBT [33–35].

## II. METHODS

### A. Crystal Growth

The MBT bulk crystals used in this work were grown by two different methods, and they can both be used for the wax-assisted exfoliation and transfer process. The crystal used to fabricate Device #1 was grown by the chemical vapor transport (CVT) method with iodine as transport agent [36]. The crystal for fabricating Device #2 was grown using the solid-state reaction method, as described in ref. [11]. All the MBT crystals used in this work were measured by X-ray diffraction (XRD) and SQUID magnetometer to ensure phase purity and high Néel temperature.

### B. The fabrication of devices

#### 1. Wax-assisted exfoliation and transfer method

Figure 1(a) illustrates the fabrication process for MBT flakes with both sides encapsulated by $AlO_x$. A small flake of MBT single crystal was first exfoliated onto a Scotch tape. Subsequently, a 3-nm-layer of aluminum was deposited onto the flake on the tape, and then oxygen was introduced into the deposition chamber to maintain a pressure of $2 \times 10^{-2}$ Pa for 5 minutes to oxidize the aluminum film.

Next, a small piece of wax (Crystalbond 509, composed of Phthalic Anhydride and Ethylene Glycol) with a typical size around 5 mm × 5 mm and thickness around 3 mm was melted onto a glass slide at 120 °C, then cooled to room temperature to form a planar surface. The tape with exfoliated MBT was gently attached to the wax, followed by gradual heating to about 50 °C. At this temperature, the wax began to infiltrate and adhere to the MBT flakes on the tape. This process must be carried out slowly to allow the wax to uniformly infiltrate the crystal surface and avoid the formation of air bubbles. Once the MBT and wax achieve sufficient adhesion, the system was cooled to room temperature, enabling the wax to solidify and bond to the crystal. The tape was then peeled away, leaving MBT flakes on the surface of wax. These samples were observed under a transmission-mode optical microscope to evaluate the thickness and size. If suitable thin flakes were found, the process proceeds to the next step. Otherwise, a fresh piece of tape was applied to the surface to exfoliate the remaining flakes. This cycle was repeated until flakes of the desired thickness are obtained, as shown in Fig. 1(b) for a 6-SL flake with $40 \times 20$ μm$^2$ that was used to fabricate Device #1. Compared to AlO$_x$- or Au-assisted exfoliation, this method has the unique advantage of allowing multiple exfoliation steps on the wax to increase the probability of finding appropriate thin flakes [37,38]. The wax also serves as a supporting layer, enabling the exfoliation of materials that are difficult to exfoliate using conventional methods. Meanwhile, the flakes on wax can be transferred to various substrates for different experimental probes.

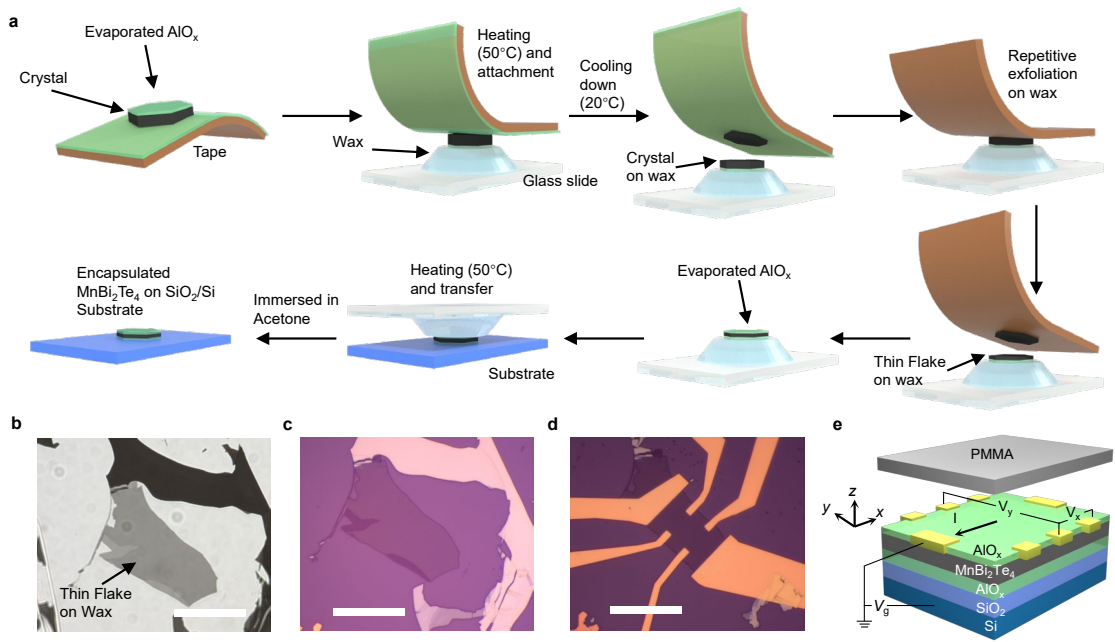

**Fig. 1.** (a) Schematic illustration of the fabrication process of MBT devices with both surfaces encapsulated by AlO$_x$. This method is developed based on a new wax-assisted exfoliation and transfer protocol. (b) Optical image of Device #1 on wax obtained by the transmission mode. [(c), (d)] Optical image of Device #1 on SiO$_2$/Si substrate before (c) and after (d) the nano-fabrication process. Scale bar: 20 μm. (e) Schematic configuration of AlO$_x$/MBT/AlO$_x$ transport device on a SiO$_2$/Si substrate.

Once a thin flake was exfoliated, the other side of the flake was also deposited a 3-nm-layer of aluminum followed by oxidization. Then the wax/AlO$_x$/MBT/AlO$_x$ structure was transferred onto a SiO$_2$/Si substrate. After the contact of wax and substrate, the transfer stage was gradually heated to 50 °C. After the full adhesion was achieved, the assembly (glass slide, wax, device and substrate) was immersed in acetone to dissolve the wax at room temperature. Figure 1(c) shows the image of AlO$_x$/MBT/AlO$_x$ on the SiO$_2$/Si substrate. All fabrication processes were conducted in an argon-filled glove box with O$_2$ and H$_2$O levels maintained below 0.1 ppm.

*2. Nano fabrication process*

After obtaining an MBT flake with both sides encapsulated by AlO$_x$, the surrounding thicker layers were removed by a needle. Then a layer of PMMA photoresist was spin-coated onto the sample, followed by baking at 60 °C for 7 minutes. Electrodes were patterned using a standard electron-beam lithography (EBL), followed by etching the AlO$_x$ on the top of the sample in an Ar ion milling machine. Metal electrodes (Cr/Au, 3/50 nm) were then deposited using a thermal evaporator connected to a glove box. Finally, the sample was immersed in acetone for the removal of photoresist films, as shown in Fig. 1(d). During the transfer between the glove box and cryostat, the devices were consistently spin-coated with a layer of PMMA. Together with the AlO$_x$ layer, it further prevents the air contamination and sample degradation. Figure 1(e) shows the final structure of the device. Acetone, PMMA, and isopropyl alcohol used in the fabrication process were all purified with molecular sieves. Throughout the entire fabrication process, the temperature is controlled below 60 °C to prevent heating-induced changes of chemical composition and degradation of crystal quality.

### C. Transport measurement

Transport measurements on Device #1 were carried out in a cryostat with the base temperature of about 1.5 K and an out-of-plane magnetic field up to 9 T. The longitudinal and Hall voltages were acquired simultaneously via SR830 lock-in amplifiers with an AC current (200 nA, 12.357 Hz) generated by a Keithley 6221 current source meter. Device #2 was measured in a dilution refrigerator with vector magnets providing a vertical field up to 9 T and a horizontal field up to 3 T. Longitudinal resistance and Hall voltages were measured using standard lock-in technique by NF5650 and NF5645. The excitation current of 10 nA at 4.56 Hz was provided by a Keithley 6221 current source. The back-gate voltage was provided by a Keithley 2400 voltage source. To correct the geometrical misalignment of the electrodes, the longitudinal and Hall signals were symmetrized and antisymmetrized with respect to the magnetic field.

## III. RESULTS

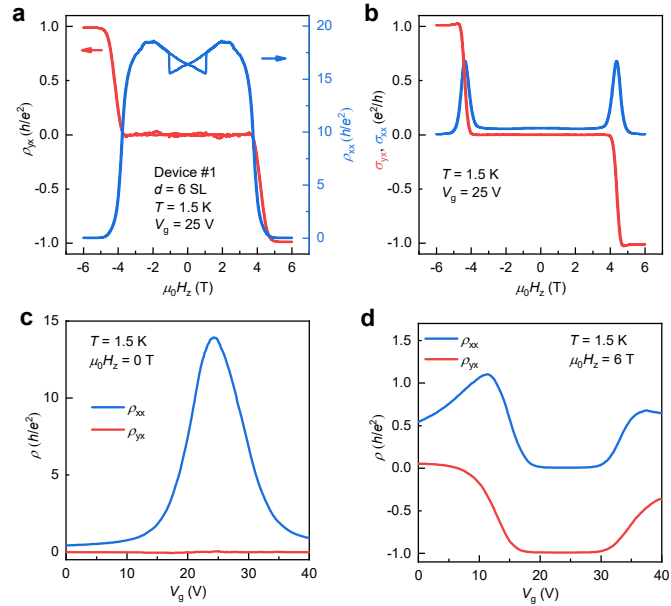

**Fig. 2.** [(a), (b)] Magnetic-field dependence of $\rho_{xx}$ and $\rho_{yx}$ (a), $\sigma_{xx}$ and $\sigma_{yx}$ (b) at $T = 1.5$ K and $V_g = 25$ V for Device #1 with $d = 6$ SL. [(c),(d)] Gate-voltage-dependent $\rho_{xx}$ and $\rho_{yx}$ in the AFM state at $\mu_0H = 0$ T (c) and the FM state at 6 T (d).

We first focus on the axion insulator state in Device #1, in which the MBT flake has a thickness $d = 6$ SL. Figure 2(a) displays the longitudinal resistivity ($\rho_{xx}$) and Hall resistivity ($\rho_{yx}$) measured at 1.5 K with a back-gate voltage $V_g = 25$ V. In the AFM state, $\rho_{xx}$ maintains a high value ($\approx 15\ h/e^2$), while the $\rho_{yx}$ forms a zero plateau within the field range $|\mu_0 H| < 3.5$ T. Figure 2(b) shows the sheet conductivity calculated by $\sigma_{xx} = \rho_{xx}/(\rho_{xx}^2 + \rho_{yx}^2)$ and $\sigma_{yx} = \rho_{yx}/(\rho_{xx}^2 + \rho_{yx}^2)$. The Hall conductivity $\sigma_{xy}$ also exhibits a very flat zero plateau in the AFM state. Figure 2(c) shows the evolution of $\rho_{xx}$ and $\rho_{yx}$ as a function of $V_g$ at $\mu_0 H_z = 0$ T. At the charge neutrality point (CNP), the bulk is insulating and the conduction of the two surfaces cancel out. All these features are typical signatures of an axion insulator with zero Chern number. Compared to the axion insulator state achieved in our previous work [15,39], the dual-surface AlO$_x$-encapsulated device exhibits a more insulating behavior in the AFM state, along with a more robust zero Hall plateau.

Next, we investigate the Chern insulator state at high magnetic fields. As depicted in Fig. 2(b), as the magnetic field increases, there is a $\sigma_{xx}$ peak around 4.4 T indicating the gap closing during the topological phase transition from the axion insulator state with Chern number $C = 0$ to the Chern insulator state with $C = -1$. At $\mu_0 H_z = 6$ T, the system enters the Chern insulator state, where $\sigma_{xx}$ drops to 0.01 $e^2/h$, accompanied by the increase of $\sigma_{yx}$ to a nearly quantized value of 1.01 $e^2/h$. Similarly, $\rho_{xx}$ is vanishing small and $\rho_{yx}$ presents the quantized plateau at $h/e^2$ (Fig. 2(b)). Figure 2(d) shows the $V_g$ dependence of $\rho_{xx}$ and $\rho_{yx}$ measured at $\mu_0 H_z = 6$ T and $T = 1.5$ K. Over the range $+18$ V $\leq V_g \leq +28$ V, a broad quantized $\rho_{yx}$ plateau and vanishing $\rho_{xx}$ are observed, indicative of dissipationless chiral edge transport at CNP. The strongly enhanced topological transport behavior, both in the axion insulator state and the Chern insulator state, directly demonstrates the critical role of AlO$_x$ in improving the magnetic properties and protecting the sample surface.

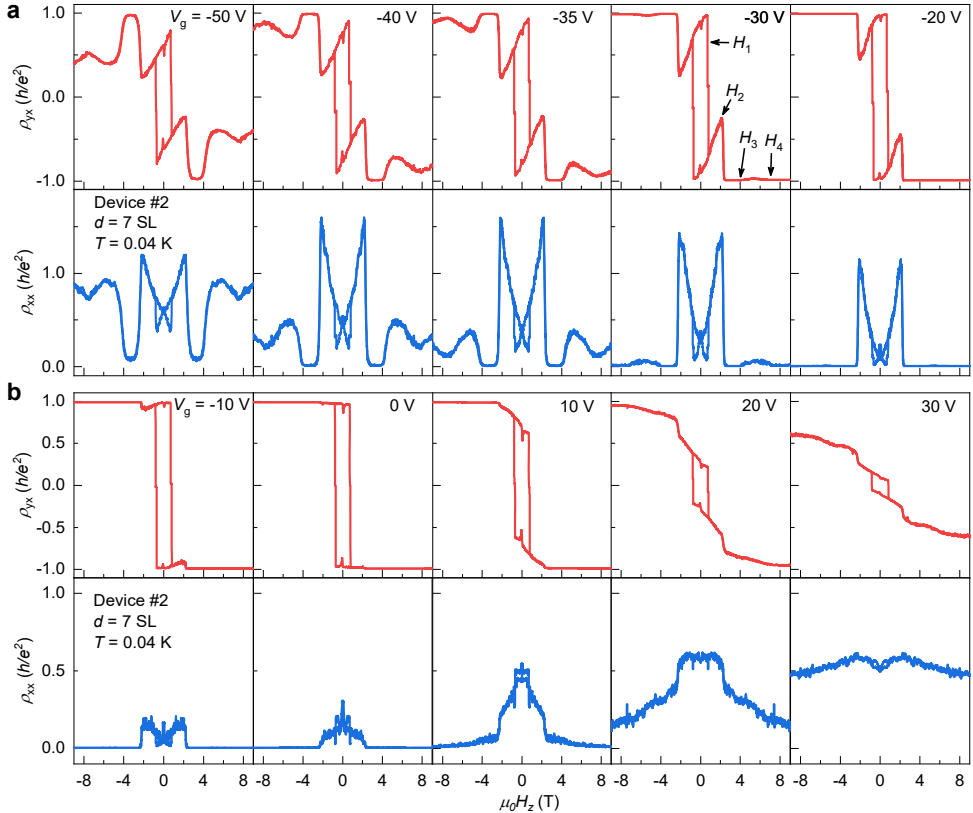

**Fig. 3.** (a) (b) $H_z$ dependence of $\rho_{xx}$ and $\rho_{yx}$ at various $V_g$s at $T = 0.04$ K for Device #2 with $d = 7$ SL. The four characteristic magnetic fields ($H_1$, $H_2$, $H_3$, $H_4$) indicate the change of magnetic configurations of MBT flakes at different $H_z$ regimes, including AFM1, AFM2 at low magnetic field with opposite Chern number, Surface Spin Flop (SSF) state between $H_2$ and $H_3$, Canted AFM (cAFM) state between $H_3$ and $H_4$, and FM state at high magnetic field.

We now examine the magnetic-field-dependent transport properties in Device #2 with thickness $d$ = 7 SL, which exhibits the QAH effect at zero magnetic field. Figures 3(a) and (b) show the $H_z$-dependent $\rho_{yx}$ and $\rho_{xx}$ loops at $T$ = 0.04 K for representative $V_g$s. At $V_g$ = -10 V, the sample exhibits the QAH effect, with $\rho_{yx}$ being nearly quantized at 0.98 $h/e^2$, and $\rho_{xx}$ dropped to 0.04 $h/e^2$ at $\mu_0 H_z$ = 0 T. The quantized $\rho_{yx}$ and vanishing $\rho_{xx}$ at CNP confirm the dissipationless nature of the chiral edge state transport in the QAH phase.

When $V_g$ deviates from the CNP, we first observe a linear behavior of $\rho_{yx}$ at low $H_z$, indicating that hole- ($V_g \leq$ -20 V) and electron-type ($V_g \geq$ 10 V) carriers begin to contributing to the ordinary Hall effect. At $V_g$ = -30 V, a cascade of quantum phase transitions emerges, characterized by the deviation and recovery of $\rho_{yx}$ quantization and zero resistance of $\rho_{xx}$ [11]. In the AFM state around zero magnetic field, $\rho_{yx}$ shows a linear behavior. Upon reaching $\mu_0 H_2 \approx$ 2.2 T, the device transits into Surface Spin Flop (SSF) state, where the subsurface SL flops to a configuration nearly parallel to the top SL [40]. This reorientation enhances near-surface magnetization which is beneficial for the QAH effect [41]. It triggers a sudden jump in $\rho_{yx}$ to the $-h/e^2$ plateau, while $\rho_{xx}$ drops from a finite value to nearly zero. The quantized state persists until $\mu_0 H_3 \approx$ 3.8 T, beyond which the device transitions into Canted AFM (cAFM) region. As the $z$-component of the bottom SL is reduced, the QAH gap is decreased, driving the device into a dissipative regime. When the magnetic field exceeds $\mu_0 H_4 \approx$ 7.8 T, the device enters the ferromagnetic (FM) state the quantized $\rho_{yx}$ plateau and vanishing $\rho_{xx}$ are restored. Notably, at $V_g$ = -40 V, the Chern insulator state only exists in the SSF state between $H_2$ and $H_3$. Even as the magnetic field increases and drives the system into the FM state, the $\rho_{yx}$ value is still away from quantized value. This behavior suggests that the QAH gap in the SSF state is larger than that in the FM state.

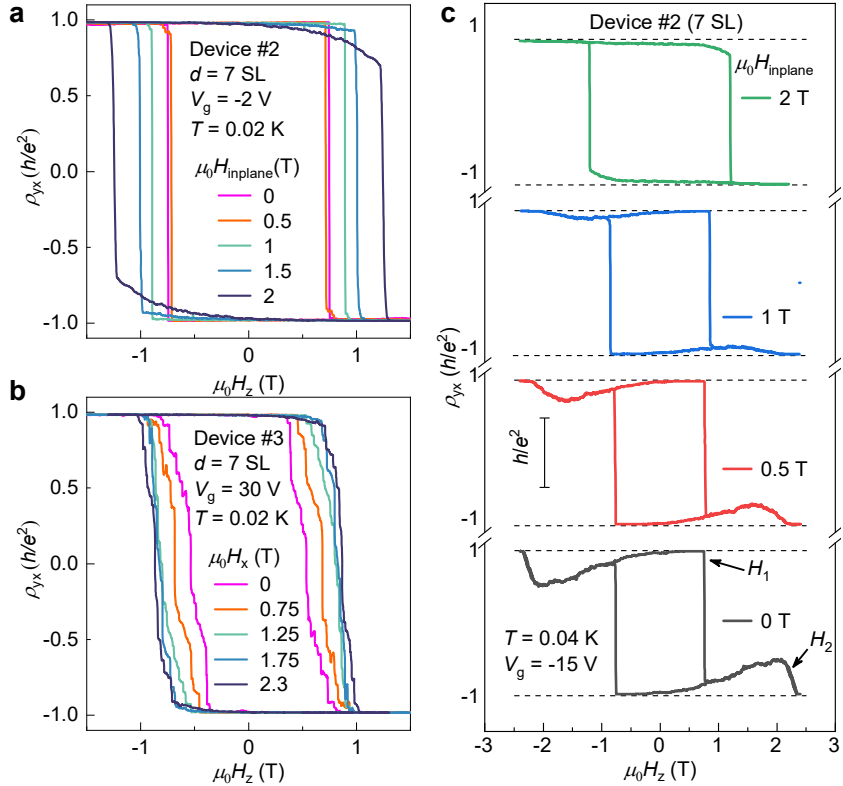

**Fig. 4.** (a) The $\mu_0 H_z$ dependent $\rho_{yx}$ loops at $T$ = 0.04 K for varied $\mu_0 H_{inplane}$ in Device #2, directly showing the enhancement of hysteresis by in-plane magnetic field. (b) The $\mu_0 H_z$ dependent $\rho_{yx}$ loops at $T$ = 0.02 K for different $\mu_0 H_x$ in Device #3 with only single-sided AlO$_x$. (c) The $\mu_0 H_z$ dependent $\rho_{yx}$ at $V_g$ = -15 V and $T$ = 0.04 K in Device #2 in different in-plane magnetic field. The QAH state between $H_1$ and $H_2$ is enhanced by the in-plane magnetic field.

Figures 4(a) and 4(b) display the $\rho_{yx}$ versus $H_z$ loops measured at CNP under selected in-plane magnetic field values for Device #2 and Device #3. Device #3 is the one used in our prior work, with only the top surface capped by AlO$_x$ layer [11]. The coercive field $H_c$, defined by the field scale when $\rho_{yx}$ = 0, is about 0.75 T for Device #2. It is much larger than $H_c$ = 0.53 T for Device #3. In classical magnetism, the value of $H_c$ is proportional to the ratio of magnetic anisotropy and saturated magnetic moment [42]. The significant increase in $H_c$ for Device #2 suggests that the AlO$_x$ at the MBT bottom surface further increases the interfacial PMA, which leads to a better square-shaped QAH loop. Notably, Fig. 4(b) displays Barkhausen jumps in the $\rho_{yx}$ curves, indicative of a multidomain structure [42], whereas Fig. 4(a) shows sharper transitions, suggesting a single-domain configuration.

In addition to the enhancement of hysteresis, the in-plane field also strengthens the AFM QAH effect between $H_1$ and $H_2$. Figure 4(c) displays the impact of $H_{inplane}$ on the $\rho_{yx}$ versus $H_z$ loops in the hole-doped region ($V_g$ = -15 V) at $T$ = 0.04 K for Device #2. In the absence of an in-plane magnetic field, the $\rho_{yx}$ exhibits linear behavior between $H_1$ and $H_2$ due to hole type charge carrier. As $H_{inplane}$ increases, $\rho_{yx}$ is evidently enhanced to form a well-developed quantum plateau and the transition at $H_2$ becomes more gradual. When the in-plane magnetic field reaches 2 T, the $\rho_{yx}$ exhibit a common behavior with $V_g$ = -2 V in Fig. 4(a). Significantly, the application of an in-plane magnetic field enlarges the QAH phase which align with our prior findings [9].

The enhancement of coercivity and QAH effect by $H_{inplane}$ contrasts starkly with that in magnetically doped TI with FM order, where an in-plane field significantly decreases the $H_c$ and suppresses the QAH [43]. The dual-surface AlO$_x$-encapsulated 7-SL MBT studied here confirm such peculiar behaviors in an even more impressive way, which demonstrate unambiguously that they are intrinsic properties of AFM QAH. Although the underlying mechanism is still unknown, we show here that they are robust features insensitive to the domain structure, and can be further enhanced when both surfaces are encapsulated by AlO$_x$. These results provide important new clues for finding the microscopic explanation for this unique behavior in MBT.

## IV. CONCLUSIONS

In summary, we developed a wax-assisted exfoliation and transfer method that enables the fabrication of high-quality MBT devices with both surfaces encapsulated by AlO$_x$. The strongly enhanced axion insulator and QAH states observed here resolve the longstanding discrepancy between layer-thickness-dependent magnetic and transport behaviors observed in experiments and those predicted by theoretical models [20,44]. The combination of wax-assisted protocol and AlO$_x$ encapsulation can also be applied to the exfoliation of high-binding-energy materials and transfer to various substrates, which will pave the way for exploring the quantum phenomena and potential applications in a wider variety of 2D materials and devices.

## V. ACKNOWLEDGMENTS

Yayu Wang was supported the Basic Science Center Project of Natural Science Foundation of China Grant No. 52388201, the Innovation program for Quantum Science and Technology Grant No. 2021ZD0302502, and the New Cornerstone Science Foundation through the New Cornerstone Investigator Program. Chang Liu was supported by fundings from National Natural Science Foundation of China Grant No. 12274453, Beijing Nova Program Grant No. 20240484574, and the Innovation program for Quantum Science and Technology Grant No. 2021ZD0302502